\documentclass[twocolumn,aps,superscriptaddress]{revtex4}
\usepackage{amssymb}
\usepackage{amsfonts}
\usepackage{amsmath}
\usepackage{amsxtra}
\usepackage{amscd}
\usepackage{bm}
\usepackage{amsthm}
\usepackage{graphicx}
\usepackage{epsf}
\usepackage{bbold}
\usepackage{xcolor}
\usepackage{mathalfa}
\usepackage{makecell} 
\usepackage{multirow}
\usepackage{hhline}
\usepackage{eucal}
\usepackage{stackengine}

\newcommand{\eq}[1]{Eq.(\ref{#1})}

\newcommand{\beq}{\begin{equation}}
\newcommand{\eeq}{\end{equation}}
\newcommand{\bmul}{\begin{multline}}
\newcommand{\emul}{{\end{multline}}}
\newcommand\beqa{\begin{eqnarray}}
\newcommand\eeqa{\end{eqnarray}}
\newcommand\bea{\begin{array}}
\newcommand\eea{\end{array}}
\newcommand\ba{\begin{array}}
\newcommand\ea{\end{array}}
\newcommand{\nn}{\nonumber}

\newcommand{\neqa}{\nonumber\end{eqnarray}}

\begin{document}

\title{Flat bands for electrons in rhombohedral graphene multilayers with a twin boundary}

\author{Aitor Garcia-Ruiz$^{1,2}$, Sergey Slizovskiy${}^{1,2}$, Vladimir I. Fal'ko}

\affiliation{National Graphene Institute, University of Manchester, Booth Street East, Manchester M13 9PL, UK}
\affiliation{Department of Physics and Astronomy, University of Manchester, Oxford Road,Manchester, M13 9PL, UK}
\affiliation{Henry Royce Institute for Advanced Materials, University of Manchester, Oxford Road, Manchester, M13 9PL, UK}
\begin{abstract} 
Topologically protected flat surface bands make thin films of rhombohedral graphite an appealing platform for searching for strongly correlated states of 2D electrons.  In this work, we study rhombohedral graphite with a twin boundary stacking fault and analyse  the semimetallic and topological properties of low-energy bands  localised at the surfaces and at the twinned interface. We derive an effective 4-band low energy model, where we implement the full set of Slonczewski-Weiss-McClure (SWMcC) parameters, and find the conditions for the bands to be localised at the twin boundary,  protected from the environment-induced disorder. This protection together with a high density of states at the charge neutrality point, in some cases -- due to a Lifshitz transition, makes this system a promising candidate for hosting strongly-correlated effects.

\end{abstract}

\maketitle
\section{Introduction}\label{Sec.I}
In the recent years, multilayer graphenes were found to host various correlated phases of matter driven by electron-electron interactions: superconductivity \cite{cao2018correlated,cao2018unconventional,Zhou2021superconductivity,park_tunable_2021,park_robust_2022}, ferromagnetism \cite{Sharpe_Emergent_2019,sharpe_evidence_2021}, nematic state \cite{Rubio-Verd2021}, and Mott insulator \cite{xu2021tunable,shen_correlated_2020,shi_electronic_2020}. The electron correlation effects in these systems are promoted by the characteristically flat low-energy bands   \cite{yin_dimensional_2019,Bistritzer_2011,slizovskiy_films_2019,mao_evidence_2020,seiler_quantum_2022,Lau_Designing_2021}. Among all these systems, few-layer rhombohedral (ABC) graphenes are the only ones which can be grown using chemical vapour deposition \cite{bouhafs_synthesis_2021} without the need to assemble twistronic structures with a high precision of crystallographic alignment. The low-energy bands in ABC films are set by topologically protected surface states, hence, it is affected by external environment. As a result, their dispersion depends both on the number of layers in the film, encapsulation and vertical electric bias, so that the ABC graphenes may behave both as compensated semimetals and gapful semiconductors \cite{slizovskiy_films_2019}.    

\begin{figure}
\begin{center}
\includegraphics[width=1\columnwidth]{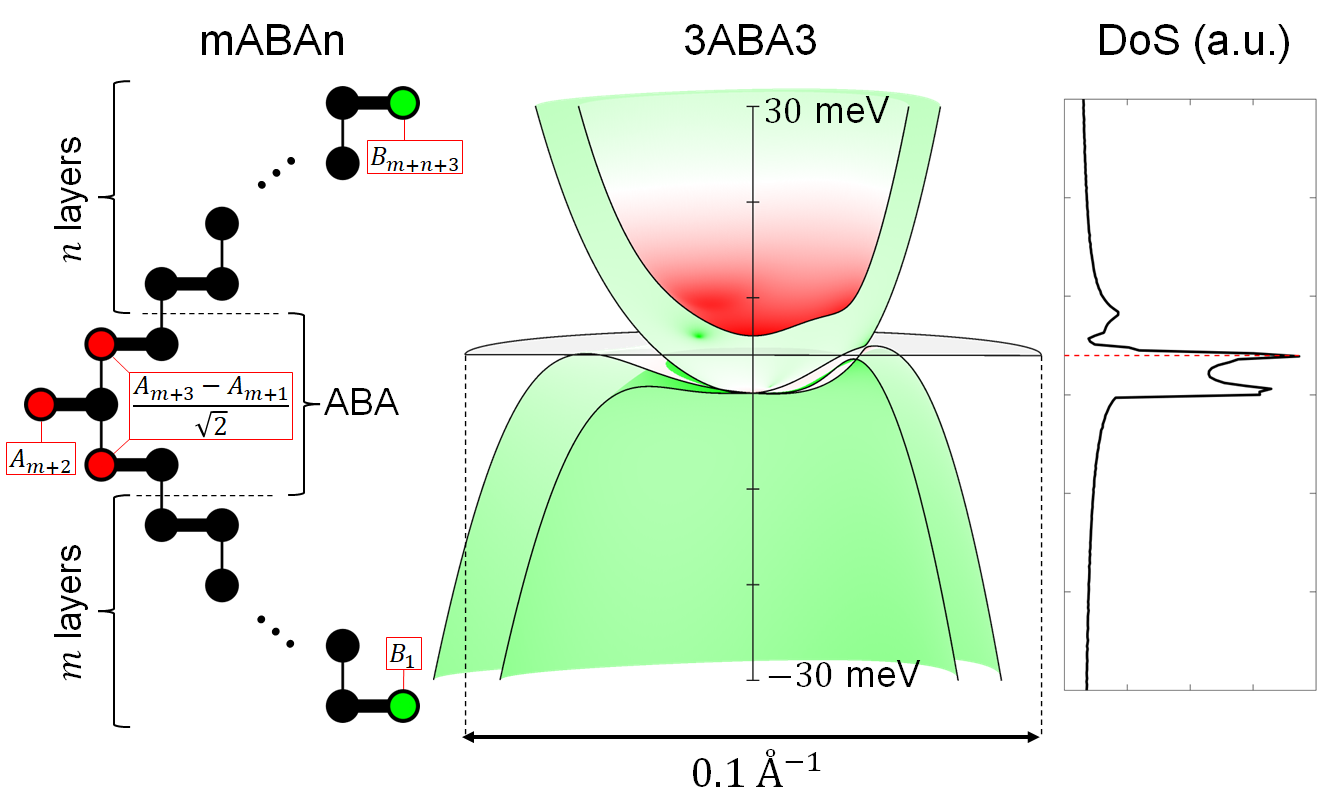}
\caption{Left: Sketch of multilayer rhombohedral graphene (mABAn) with one twin boundary (ABA 'trilayer'). The low-energy basis is highlighted in red/green for the relevant twin boundary sites/surface layers. Middle: Low-energy band structure of a 3ABA3 film across energy window $\pm30$ meV, with the colour coding of bands according to the dominant location of their wave function. Right: Density of states (DoS) of electrons in a 3ABA3 multilayer with the Fermi level in an undoped structure coinciding with the van Hove singularity.
\label{fig:Twin_Boundary}}
\end{center}
\end{figure}

A rhombohedral graphitic film with one stacking fault such as as twin boundary, Fig. \ref{fig:Twin_Boundary}, also host low-energy flat bands \cite{Arovas2008,Muten_2021}: four rather than two specific for ABC graphene. The additional two bands come from the twin boundary inside the film, hence, they can be protected from the environmental influences due to screening by the surface states. Here, we study the low-energy spectra of thin films of twinned ABC graphenes {with $N = m+n+3$ layers} such as 'mABAn' multilayer  sketched in Fig. \ref{fig:Twin_Boundary}, where the twin boundary appears as a Bernal (ABA) trilayer buried inside the film with $n$ and $m$ rhombohedrally stacked (ABC and CBA) layers above and underneath it. In Fig. \ref{fig:Twin_Boundary} we also present four low-energy bands in a 9-layer film (3ABA3) with a twin boundary at the middle layer, which illustrates that such systems are semimetals and that - in some of these systems - there might be at least one low energy band located at the twinned interface. Moreover, we notice that a neutral (undoped) 3ABA3 multilayer has an additional feature: the electron Fermi energy in it is close to the Lifshitz transition \cite{Lifshitz_Anomalies_1960,Fermi_Volovik_1994,Varlet_Anomalous_2014}, marked by the van Hove singularity in the density of states, Fig. \ref{fig:Twin_Boundary}. 

The presented-below analysis of band structure of twinned multilayers of ABC graphene is based on the hybrid $\bm{ k\cdot p}$ - tight binding theory which accounts for the full set of Sloczweski-Weiss-McClure (SWMcC) parameters for graphite \cite{Slonczewsi1958,McClure1957,McClure1960}, in section \ref{Sec.II}.  Taking all SWMcC parameters into account appear to be important, as (similarly to what has been found in monolithic ABC films \cite{slizovskiy_films_2019}) the next-neighbour/layer hoppings and coordination-dependent on-carbon potentials lift an artificial degeneracy of band edges predicted by the minimal model accounting for only closest neighbour hopping \cite{Muten_2021,Koshino_trigonal_2009}. In Sec. \ref{Sec.III}, we develop and test an effective 4-band model for rhombohedral structures with one twin boundary, which improves the low-energy Hamiltonian derived in Ref. \cite{Muten_2021}, and use it to study the Berry curvature and the magnetic moment of the bands, Sec. \ref{Sec.IV}. {This effective Hamiltonian could provide
an analytical tool for further studies of correlation effects. 
}

\section{SWMcC model for multilayers with various stackings}\label{Sec.II}
 In the basis of sublattice amplitudes for electron states in a mABAn {N-layer films ($N=n+m+3$)}, ${\Psi}^{\dagger}=\left(
\begin{matrix}\psi_{A_1},&\psi_{B_1},&\cdots,& \psi_{A_N},
\psi_{B_{N}}\end{matrix}\right)^{\dagger}$, the Hamiltonian, which will be used to describe the subbands in it,  is written as
\beqa \label{Eq:Ham}
\mathcal{H}&=&
\left(
\begin{matrix}
H_\mathrm{g}^s&V&W&\cdots&0&\cdots&0&0&0\\
V^\dagger&H_\mathrm{g}^b&V&\cdots&0&\cdots&0&0&0\\
W^\dagger&V^\dagger&H^b_\mathrm{g}&\cdots&0&\cdots&0&0&0\\
\vdots&\vdots&\vdots&\ddots&\vdots&
\vdots&\vdots&\vdots&\vdots\\
0&0&0&\cdots&\mathcal{H}_{\mathrm{ABA}}&\cdots&0&0&0\\
\vdots&\vdots&\vdots&\cdots&\vdots&\ddots&\vdots&\vdots&\vdots\\
0&0&0&\cdots&0&\cdots&H^b_\mathrm{g}&V^\dagger&W^\dagger\\
0&0&0&\cdots&0&\cdots&V&H^b_\mathrm{g}&V^\dagger\\
0&0&0&\cdots&0&\cdots&W&V&H_\mathrm{g}^s
\end{matrix}
\right), \nn \\
&&\mathcal{H}_{ABA}=
\left(
\begin{matrix}
H^b_\mathrm{g}&V&\tilde{W}\\
V^{\dagger}&H^b_\mathrm{g}-\sigma_z\Delta'&V^\dagger\\
\tilde{W}^\dagger&V&H^b_{\mathrm{g}}
\end{matrix}
\right), \\
&&H_\mathrm{g}^s=
H_{\mathrm{g}}+
\left(
\begin{matrix}
\Delta'&0\\
0&0
\end{matrix}
\right)+\Delta_s\hat{\mathbb 1}_2,
\quad
H^b_\mathrm{g}=H_{\mathrm{g}}+
\Delta'\hat{\mathbb 1}_2
,\,\nonumber
\\
&&H_g=v
\left(
\begin{matrix}
0&\pi_\xi^*\\
\pi_\xi&0
\end{matrix}
\right),\quad
V=
\left(
\begin{matrix}
-v_4\pi_\xi&\gamma_1\\
-v_3\pi_\xi^*&-v_4\pi_\xi
\end{matrix}
\right),
\nonumber\\
&&
W=
\left(
\begin{matrix}
0&0\\
\gamma_2/2&0
\end{matrix}
\right)\quad
\tilde{W}=
\left(
\begin{matrix}
\gamma_5/2&0\\
0&\gamma_2/2
\end{matrix}
\right).\nonumber
\eeqa
Here, ${\hat{\mathbb 1}}_2$ is a $2\times2$ unit matrix, $\pi_\xi\equiv\xi p_x+ip_y$, with $\boldsymbol{p}=(p_x,p_y)$ being the valley momentum measured from $\hbar\boldsymbol{K}_\xi=\hbar\xi
\frac{4\pi}{3a}(1,0)$. 
{
$H_{\mathrm{g}}$ and $H_{\mathrm{g}}^{b/s}$ are Hamiltonians of free-standing graphene and graphene inside/ at the surface of the structure, respectively. 
Matrices $V$ and $W$  describe the nearest and next-nearest layer couplings, and they are assumed to be independent of the distance to the surface layers.} Below we use the following values of parameters implemented in \eq{Eq:Ham}:  $v = 1.02 \cdot10^6$ m/s, $v_3 = 0.102 \cdot10^6$ m/s, $v_4= 0.022\cdot10^6$ m/s, $\gamma_1=390$ meV, $\Delta'=25$ meV, $\gamma_2=-17$ meV, $\gamma_5=38$ meV \cite{yin_dimensional_2019, Ge2021Control}. In addition, we account for energy shift, $\Delta_s$, of the surface orbitals which captures the influence of the encapsulation and other environmental conditions.

\begin{figure*}
\begin{center}
\includegraphics[width=2\columnwidth]{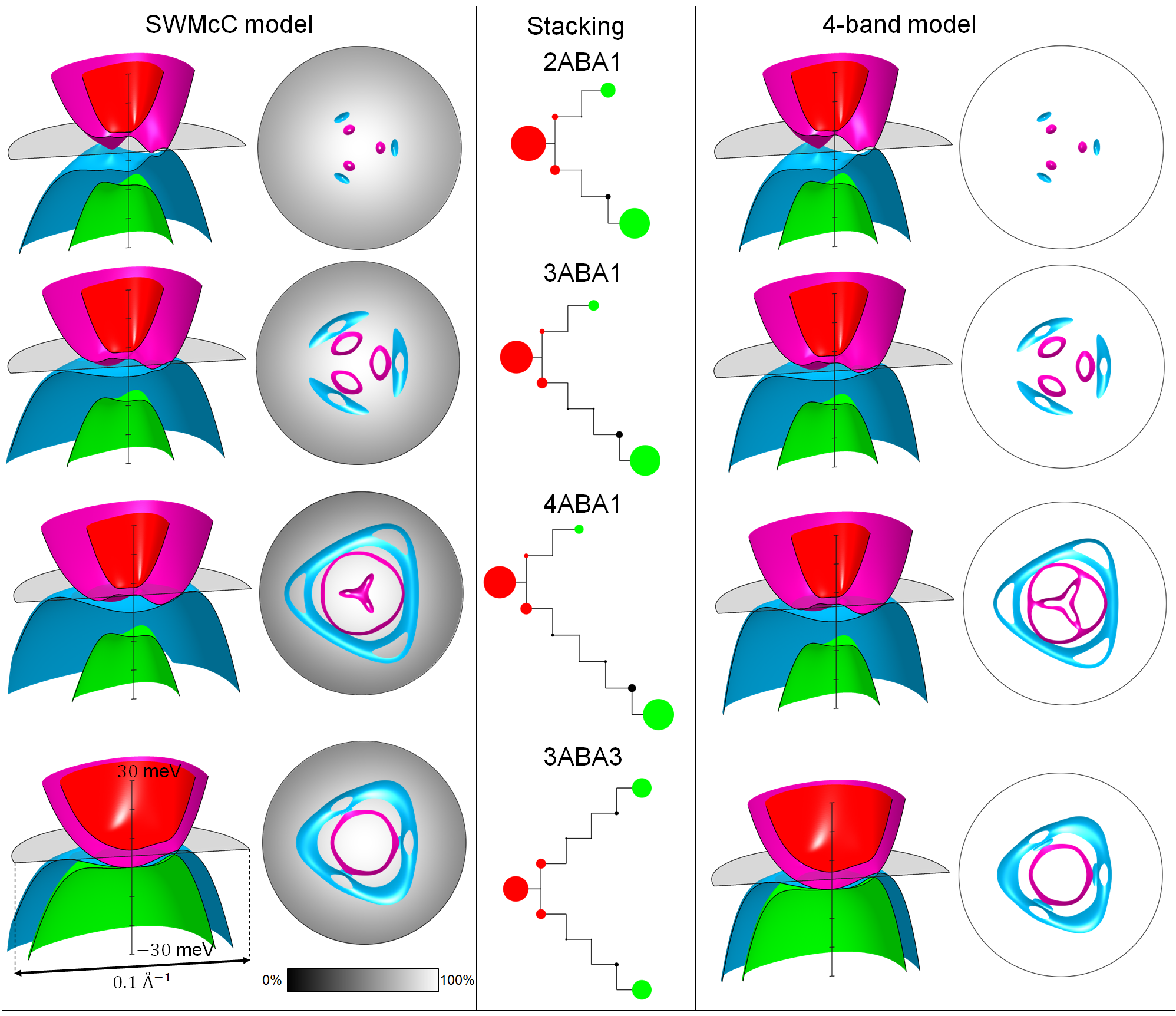}
\caption{\label{fig:Band_structures} Comparison between the full SWMcC (left) and effective 4-band model (right) for the low-energy band dispersion's for various nABAm films (middle) within $\pm30$ meV near the Fermi level in undoped structures; the Fermi-energy contours are plotted with a broadening of $\pm 1$ meV. The background behind the Fermi contours indicates the total probability to find the carriers of the 4 low-energy bands on the 4 low energy orbitals: this demonstrates that the range of validity of 4-band model extends over the range of momentum space fully accommodating the Fermi energy cuts of the bands. {In the middle panel, the sizes of the circles representing the sublattices (in red, green and black, for the twin boundary, surface and bulk orbitals, respectively) were computed by integrating the amplitude squared of the wavefunction over all bands in the energy window shown on the dispersion plot.}}
\end{center}
\end{figure*}

\begin{figure}
\begin{center}
\includegraphics[width=1\columnwidth]{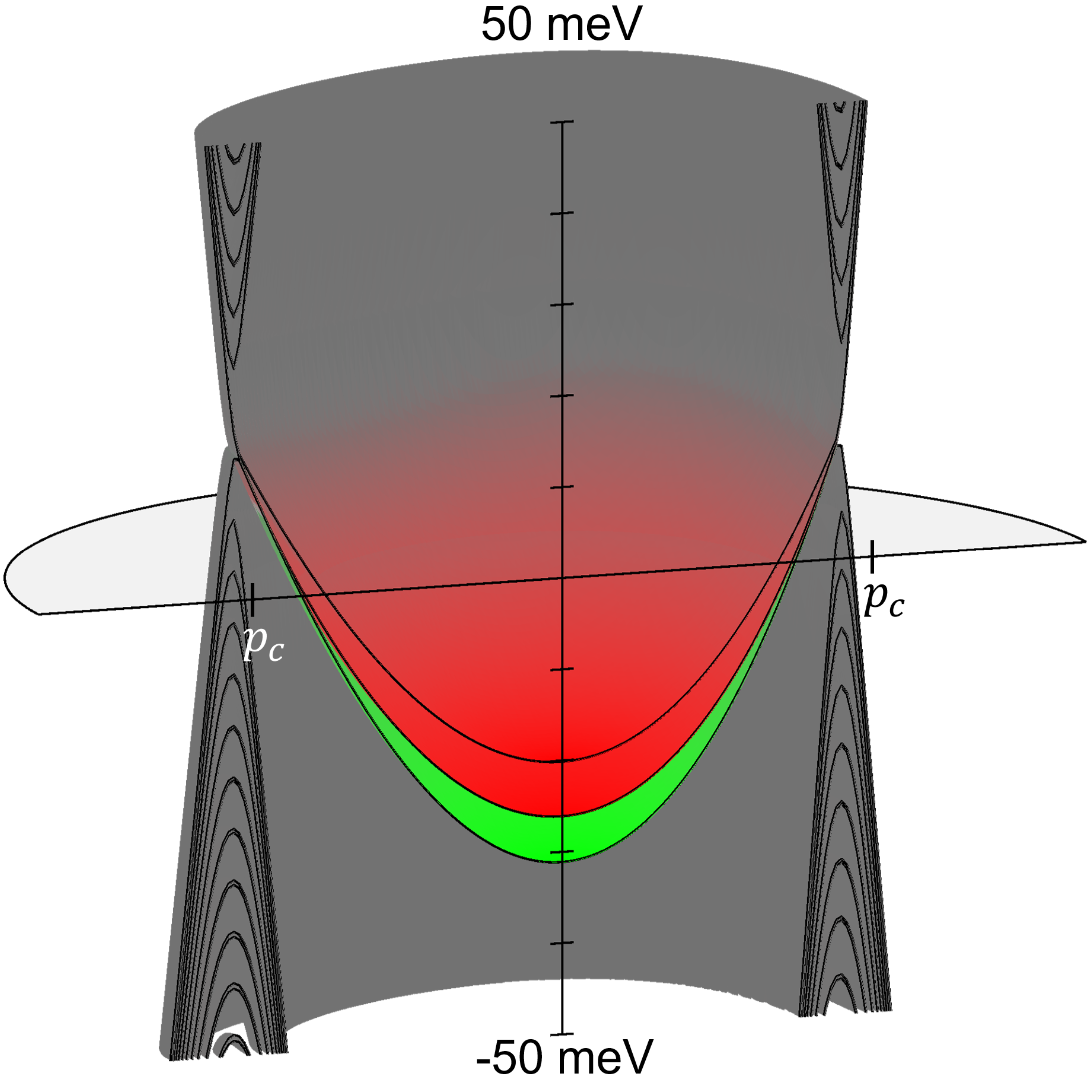}
\caption{\label{fig:largem} {Dispersion for a 200ABA200 structure, showing the twin boundary bands in red, surface bands in green and bulk bands in grey color.  We used a non-zero value of surface energy $\Delta_s = -5 \,\rm meV$ to visually separate the bands. 
}}
\end{center}
\end{figure}

\begin{figure*}
\begin{center}
\includegraphics[width=2\columnwidth]{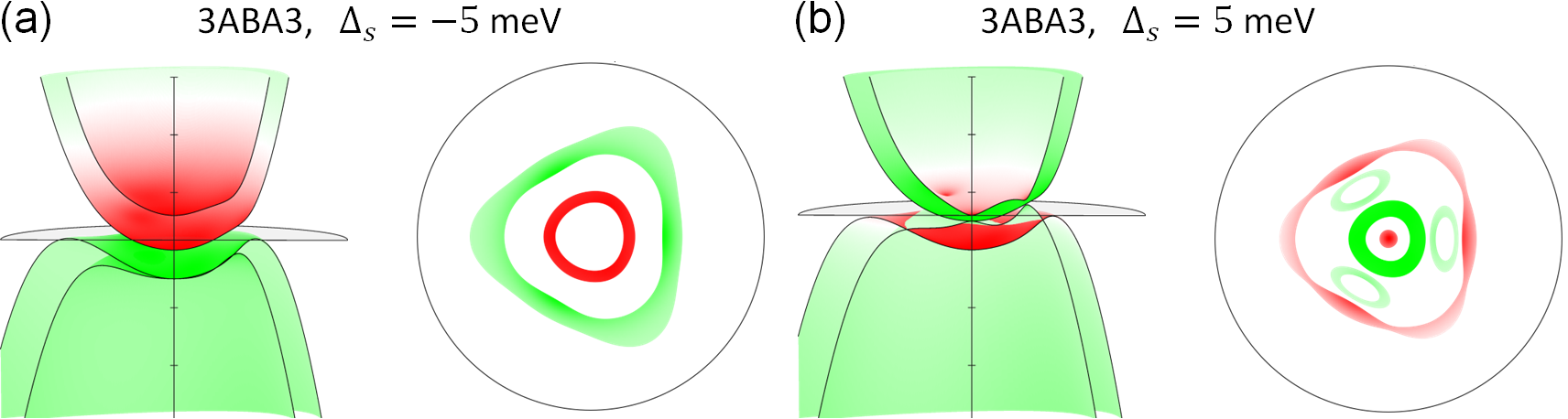}
\caption{\label{fig:Effect_of_DeltaS}  The influence of the surface orbitals energy shift, $\Delta_s$, on on the distribution of wave function weight between the surface (green) and twin boundary (red) for a 3ABA3 film. Images in this row should be compared to the spectra and wave function weights displayed in Fig.\ref{fig:Twin_Boundary} for $\Delta_s = 0$}.
\end{center}
\end{figure*}

In the Hamiltonian for the band energies around the centre of $\boldsymbol{K}_\pm $ valleys,  parameter $\gamma_1$ sets the largest energy scale.  As a consequence, in rhombohedral graphite with a twin boundary, we observe a clear spectral separation of the bands in the dispersion, where a set of $m+n+1$ conduction (valence) bands are split by $\pm \gamma_1$  from the two isolated pairs of conduction and valence bands with dispersions illustrated  in Fig. \ref{fig:Band_structures} for several exemplary multilayers.
{The bulk (split) band edges  appear at \cite{AitorRaman,shi_electronic_2020,slizovskiy_films_2019}  $\pm \frac{\pi \gamma_1}{\max(m,n)} \approx \pm \frac{1 \, {\rm eV}}{\max(m,n)}$ near the Fermi level at $|p| \approx p_c \equiv \gamma_1/v \approx 0.058\, \AA^{-1}\hbar$, as in Fig.\ref{fig:largem}. 
For $m,n \lesssim 50$ the separation of bulk states from interface and surface states  exceeds the low energy dispersion, $p_c^2/(2 m_e) \approx 40 \, \rm meV$. In this case,  the bulk remains insulating, and the effective theory for low-energy bands, presented below, applies in full.  

In the formal `bulk limit',  $m,n \gg 50$, where the bulk modes cross the Fermi level, 
the surface and the twin boundary  states remain the same for $|p|<p_c$, with similar parabolic dispersions,
$E_\text{surface 1,2} = \frac{ p^2}{2 m_e} +\Delta_s$,  \
$E_{\psi_a} =  \Delta' -\frac{\gamma_5}{2}  + \frac{ p^2}{2 m_e}$,\ \ $ E_{\psi_{B,m+2}} = \frac{p^2}{2 m_{e}'}$ (see Fig.\ref{fig:largem}),   where  $\psi_a=(\psi_{A_{m+3}}-\psi_{A_{m+1}})/\sqrt{2}$, $m_e=
\left(
\frac{4 v v_4}{\gamma_1}+
\frac{2 v^2\Delta'}{\gamma_1^2}
\right)^{-1}$, and $
m_e'=
\left(
\frac{4 vv_4}{\gamma_1}+
\frac{v^2(\Delta'+\gamma_5/2)}{\gamma_1^2}
\right)^{-1}.$ However, close to momentum $p_c$, the interfacial states blend into the bulk spectrum. Also, Fermi level may shift due to a charge redistribution between the bulk and the interface states. 
We do not consider this case in detail since in the recent experimental studies it is  hard to find rhombohedral crystals  with $>$50 ABC layers and without stacking faults \cite{Nery2021}. }

\section{Effective 4-band model for twinned rhombohedral films}\label{Sec.III}
To study the low-energy dispersion, we employ degenerate perturbation theory \cite{Min_2008}, which allows us to construct an effective $4\times4$ Hamiltonian out of the low-energy basis, highlighted in red in Fig. \ref{fig:Twin_Boundary}. This basis consists of three sublattice amplitudes of the non-dimer orbitals, $\psi_{B_1},\psi_{A_{m+2}}$ and $\psi_{B_{N}}$, and the antisymmetric combination of sublattice amplitude of orbitals dimerised with the layer at the twin boundary, $\psi_a=(\psi_{A_{m+3}}-\psi_{A_{m+1}})/\sqrt{2}$. In turn, the high-energy basis encompasses the rest of $p_z$ orbitals the dimer sites and a symmetric orbital $\psi_s=(\psi_{A_{m+3}}+\psi_{A_{m+1}})/\sqrt{2}$. 
The matrix elements of the low-energy effective Hamiltonian can be determined from the degenerate perturbation theory \cite{Min_2008} around the valley $\hbar\boldsymbol{K}_\xi$ point, 
\begin{equation}
\label{Eq:MatrixElements}
\langle\psi_i|\mathcal{H}_{\mathrm{eff}} |\psi_j\rangle=
\langle\psi_i|H_{g}\left\{Q[-H_{\perp,\mathrm{X}}^{-1}]QH_g\right\}^{n-1}|\psi_j\rangle,
\end{equation}
where $H_{\perp,\mathrm{X}}$ ($\mathrm{X} = \mathrm{R} \text{ or } \mathrm{T}$) is the high energy Hamiltonian acting on the high-energy basis of B-A dimer bonds inside the two rhombohedral stacks,
\begin{align}
H_{\perp,\mathrm{R}}^{-1}
\approx
\left(
\begin{matrix}
-\Delta'/\gamma_1^2&
1/\gamma_1\\
1/\gamma_1&
-\Delta'/\gamma_1^2\\
\end{matrix}
\right),
\end{align}
or between the high-energy basis around the twin-boundary, $\left\{\psi_{B_{m+2}},\psi_s\right\}$,
\begin{align}
H_{\perp,\mathrm{T}}^{-1}
\approx
\left(
\begin{matrix}
-(2\Delta'+\gamma_5)/(4\gamma_1^2)&
1/(\sqrt{2}\gamma_1)\\
1/(\sqrt{2}\gamma_1)&
-\Delta'/\gamma_1^2\\
\end{matrix}
\right),
\end{align}
in the high-energy basis adjacent to the twin-boundary. In Eq. (\ref{Eq:MatrixElements}), $Q$ is a  projector onto the subspace spanned by the low-energy basis. 

Then, for $n,m>0$, the low-energy Hamiltonian, written in  the basis of $\left\{\psi_{B_1},\psi_a,\psi_{B_{N}},\psi_{A_{m+2}}\right\}$, takes the form
\begin{widetext}
\begin{align}\label{Eq:Eff_Ham}
\mathcal{H}_{\mathrm{eff}}=&
\left(
\begin{matrix}
\frac{ p^2}{2 m_e} +\Delta_s& \frac{-\gamma_1 \mathcal{X}_{m+1}}{\sqrt{2}} & 0 & \frac{-\gamma_1 \mathcal{X}_{m+2}- v_3 \pi_\xi \mathcal{X}_{m}+\gamma_2 \mathcal{X}_{m-1}/2}{2}
\\
\frac{-\gamma_1 \mathcal{X}^*_{m+1}}{\sqrt{2}}  &    \Delta' -\frac{\gamma_5}{2}  + \frac{ p^2}{2 m_e}   &  \frac{\gamma_1 \mathcal{X}^*_{n+1}}{\sqrt{2}} &  0 \\
  0 &   \frac{\gamma_1 \mathcal{X}_{n+1}}{\sqrt{2}} & \frac{ p^2}{2 m_e}+\Delta_s &  \frac{-\gamma_1 \mathcal{X}_{n+2}- v_3 \pi_\xi \mathcal{X}_{n}+\gamma_2 \mathcal{X}_{n-1}/2}{2} \\
\frac{-\gamma_1 \mathcal{X}^*_{m+2}- v_3 \pi_\xi^* \mathcal{X}^*_{m}+\gamma_2 \mathcal{X}^*_{m-1}/2}{2} & 0 &   \frac{-\gamma_1 \mathcal{X}^*_{n+2}- v_3 \pi_\xi^* \mathcal{X}^*_{n}+\gamma_2 \mathcal{X}^*_{n-1}/2}{2} &  \frac{p^2}{2 m_{e}'} 
\end{matrix}
\right),&\\
&\mathcal{X}_n=
\sum_{n_1,n_2,n_3}
\frac{(n_1+n_2+n_3)!}{n_1!n_2!n_3!}
\left(
-\frac{v\pi_\xi^*}{\gamma_1}
\right)^{n_1}
\left(
\frac{v_3\pi_\xi}{\gamma_1}
\right)^{n_2}
\left( -\frac{\gamma_2}{2\gamma_1} \right)^{n_3},
\nonumber \,
\mathcal{X}_0 = 1, \,  \mathcal{X}_{n<0} = 0,&
\end{align}
\end{widetext}
where $p\equiv|\pi_\xi|$, the sum in $\mathcal{X}_n$ is extended to all positive integers $n_1,n_2$, and $n_3$ satisfying $n_1+2n_2+3n_3=n$,  $m_e=
\left(
\frac{4 v v_4}{\gamma_1}+
\frac{2 v^2\Delta'}{\gamma_1^2}
\right)^{-1}$, and $
m_e'=
\left(
\frac{4 vv_4}{\gamma_1}+
\frac{v^2(\Delta'+\gamma_5/2)}{\gamma_1^2}
\right)^{-1}$. For a special case of $m=0$ (and/or $n=0$), we substitute $m_e$ in the on-site energy of $\psi_{B_1}$ (and/or $\psi_{B_{N}}$) with $2 m_e'$, and, for the case of ABA trilayer, $m=n=0$, there is a direct $\gamma_2/2$ hopping between $B_1$ and $B_3$.

In Fig. \ref{fig:Band_structures}, we present a comparison between band structures obtained by diagonalising the full Hamiltonians (\ref{Eq:Ham}) and (\ref{Eq:Eff_Ham}) for four different stacking configurations. The two sets of low-energy dispersions almost coincide, as well as, the Fermi contours at charge neutrality. We also present the Fermi contour at charge neutrality (broadened by $\pm 0.5$ meV), to highlight the existence of both electron and holes pockets. Interestingly, for 4ABA1 and 3ABA3 films, the Fermi level lies very close to the Lifshitz transition point, which enhances the density of states. To estimate the range of validity of the low-energy model in \eq{Eq:Eff_Ham}, we use the second column of Fig. \ref{fig:Band_structures} to indicate with the background color the sum of the squared amplitudes of the eigenvectors of the 4 low-energy bands on the low-energy orbitals, computed using the full SWMcC model. We note that the electron states near the Fermi surface at charge neutrality are located mostly on the low-energy orbitals, so that the effective model is applicable.  
\begin{figure*}
\begin{center}
\includegraphics[width=2\columnwidth]{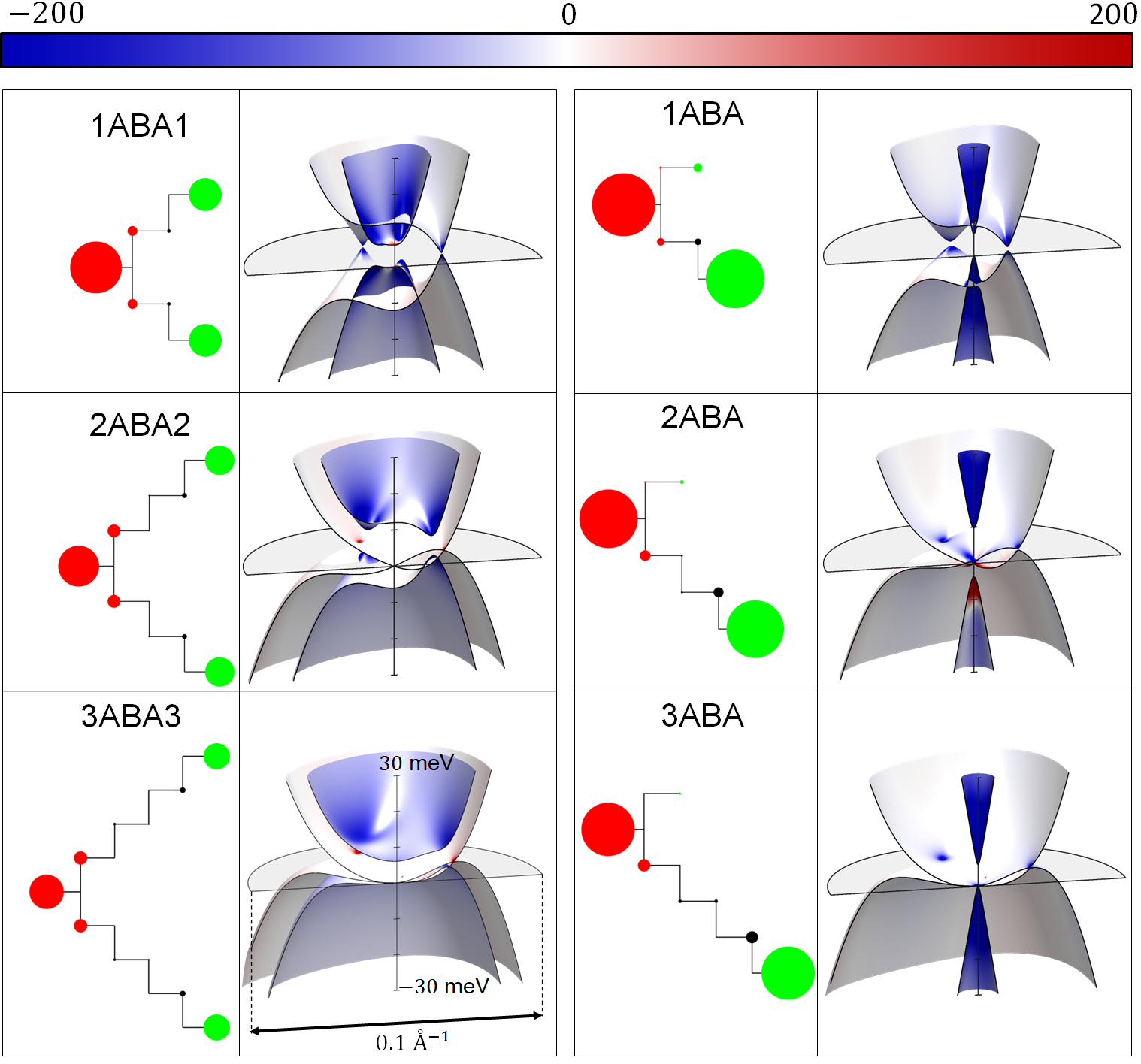}
\caption{{Low-energy bands around the $\boldsymbol{K}_+$ valley of nABAn and nABA films within an energy window of $60$ meV centred at Fermi level. Colours indicate the sign and magnitude of the $g_v$-factor. The `Dirac points' $\boldsymbol{p}_D$ (where the distance between the conduction and valence bands is minimal) are the hotspots of Berry curvature.} 
\label{fig:Magnetic_Moment}}
\end{center}
\end{figure*}

When  $m=n$, the structure of the film is mirror-symmetric with respect to the twin boundary, so that the low energy Hamiltonian decouples into two $2\times2$ blocks in the basis $\left\{\frac{\psi_{B_1} + \psi_{B_N}}{\sqrt{2}}, \psi_{A_{n+2}}, \frac{\psi_{B_1} - \psi_{B_N}}{\sqrt{2}},\psi_a \right \}$ of mirror-symmetric (s) and anti-symmetric (as) bands,
\begin{widetext}
\begin{subequations}\label{Eq:Hs_Has}
\begin{align}
& {\mathcal  H}_{m=n} = \left( \begin{matrix}
  H_{\mathrm{s}} & 0 \\
  0 &  H_{\mathrm{as}}
  \end{matrix} \right)\nonumber \\
& {H_\mathrm{s}} = \left( \begin{matrix}
  \frac{p^2}{2 m_e} +\Delta_s+ \left[ \frac{\gamma_2}{2} + \frac{p^2}{4 m_e'} - \frac{p^2}{2 m_e} \right]   \delta_{n,0}&   \frac{-\gamma_1 \mathcal{X}_{n+2} - v_3 \pi_\xi \mathcal{X}_{n}+\gamma_2 \mathcal{X}_{n-1}/2}{\sqrt{2}} \\
  \frac{-\gamma_1 \mathcal{X}^*_{n+2} -  v_3 \pi_\xi^* \mathcal{X}^*_{n}+\gamma_2 \mathcal{X}^*_{n-1}/2}{\sqrt{2}} &   \frac{p^2}{2 m'_e} 
  \end{matrix} \right)\\
& {H_{\mathrm{as}}} = 
    \left( \begin{matrix}
  \frac{p^2}{2 m_e}+
  \left[
  \frac{p^2}{4m_e'}-
  \frac{p^2}{2m_e}-
  \frac{\gamma_2}{2}
  \right]\delta_{n,0} &   -\gamma_1 \mathcal{X}_{n+1}  \\
   -\gamma_1 \mathcal{X}^*_{n+1} & \Delta' - \frac{\gamma_5}{2} +  \frac{p^2}{2 m_e}
  \end{matrix}\right) .
\end{align}
\end{subequations}
\end{widetext}

The two expressions above resemble the effective Hamiltonian for films of rhombohedral graphite \cite{Koshino_trigonal_2009} with $n+2$ and $n+1$  of layers, respectively. Following the notation in \cite{slizovskiy_films_2019}, they can be expressed as 
\beq
H_\beta=
\hat{1}_2
\varepsilon_{\beta}
(\boldsymbol{p})+
\boldsymbol{\sigma}\cdot
  \boldsymbol{d}_{\beta}
  (\boldsymbol{p}),
\eeq
where $\beta=\mathrm{s},\mathrm{as}$, $\boldsymbol{\sigma} = (\sigma_x,\sigma_y,\sigma_z)$ is the vector of Pauli matrices and the energy spectrum of conduction/valence bands is given by $\varepsilon_\beta (\boldsymbol{p}) \pm | \boldsymbol{d}_\beta (\boldsymbol{p})|$.  Similarly to thin films of rhombohedral graphite \cite{Koshino_trigonal_2009}, the band structure, shown in left column of Fig. \ref{fig:Magnetic_Moment}, 
exhibits triads of Dirac-like points located at $\boldsymbol{p}_D^{(j)}=\mathcal{R}_{\frac{2\pi }{3}j}(p_D,0)$, with $\mathcal{R}_\theta$ being the rotation operator by an angle of $\theta$. The possible values of $p_D$ are the roots of the polynomials in the off-diagonal elements of the matrices $H_{\mathrm{s}}$ and $H_{\mathrm{as}}$,  and these Dirac points are weakly gapped by 
\begin{subequations}\label{Eq:Gaps}
\begin{align}
\delta_{\rm s} = 2 \left|d_{\mathrm{s},z}(\boldsymbol{p}_D)\right|=&
\left|\frac{p_D^2v^2}{4\gamma_1^2}(2\Delta'-\gamma_5)+\Delta_s \right. \nn \\
&\left. +
\left[\frac{\gamma_2}{2}-\frac{v^2p_D^2}{4\gamma_1^2}
\left (4\gamma_1\frac{v}{v_4}+\gamma_5\right)\right] 
\delta_{n,0} \right|\nonumber\\
\delta_{\rm as} = 2 \left | d_{\mathrm{as},z}(\boldsymbol{p}_D) \right| =&
\left|\frac{\gamma_5}{2}-\Delta' \right.\nn \\
&\left. -\left[\frac{\gamma_2}{2}+\frac{v^2p_D^2}{4\gamma_1^2}
\left(4\gamma_1\frac{v}{v_4}+\gamma_5\right)\right]\delta_{n,0}\right|.\nonumber
\end{align}
\end{subequations}

To mention, for $\Delta_s=0$,  the gaps  for symmetric bands vanish, $\delta_s = 0$,  due to the imposed a degeneracy of non-dimer orbitals on the twinned interface. The environment, such as encapsulation, would influence the energies of surface orbitals, lifting the mentioned degeneracy and introducing the energy separation of surface and twin boundary bands. Low-energy band dispersions for $\Delta_s\neq 0$ are shown in the bottom panels of Fig. \ref{fig:Band_structures}, where we 
also indicate by colours  the bands which are localized  on the surface or on the twin boundary.  

\section{Berry curvature and topological valley g-factors of the low-energy bands}\label{Sec.IV}
The use of the effective 4-band model is convenient to  study the topological characteristics of rhombohedral graphite with a twin-boundary stacking fault. This includes Berry curvature of the bands and the associated magnetic moments, described in the literature in terms of valley g-factors. Large magnetic moment  affects the transport measurements, quantum dot spectra \cite{Angelika2018,Angelika2021,AngelikaBLGmoment,Ge2021Control}, valley-polarised currents \cite{Niu_valley_2007}, or an anomalous contribution to the Hall conductivity \cite{slizovskiy_films_2019}. Such an analysis is particularly easy to perform for mirror-symmetric nABAn films, where the $4\times 4$ Hamiltonian is reduced to a  pair of effective $2\times2$ Hamiltonians, one for mirror-symmetric (s) and the other for mirror-antisymmetric (as) bands. In this case the topological valley g-factor due to the Berry curvature  of conduction (+) and valence (-) bands read \cite{slizovskiy_films_2019} 
\beq\label{Eq:BerryCurvature}
    g_{v\, \beta}^\pm = 2 \frac{\Omega^+_\beta}{d} \ ; \ \Omega^{\pm}_{\mathrm{\beta}} = \mp \frac{\hbar^2}{2 d^3} {\bf d}_{\mathrm{\beta}} \cdot [\partial_{p_x} {\bf d}_{\mathrm{\beta}} \times \partial_{p_y} {\bf d}_{\mathrm{\beta}}].
\eeq    
Note that this magnetic moment and Berry curvature have opposite signs in the $\boldsymbol{K}_\pm$ valleys, so that $g_v$ would directly quantify the valley splitting in out-of-plane magnetic field.   

The band structures around $\boldsymbol{K}_+$ of three mirror-symmetric films, coloured according to valley g-factor, are presented on the left-hand side panels of Fig. \ref{fig:Magnetic_Moment}, for $\Delta_s=0$. On the right-hand side panels of Fig. \ref{fig:Magnetic_Moment}, we plot the valley g-factor for several  mirror-asymmetric structures, computed with the Hamiltonian in \eq{Eq:Eff_Ham}. 
 The topological features in these two pairs of bands are concentrated near the band edges, with valley g-factor reaching  $g_v\sim100$ at the hot-spots near Dirac points.
For the structures with a small number of layers, there are well-articulated Dirac points with large orbital magnetic moment concentrated at them and $g_v \sim 10^3 $, like observed earlier in ABA graphene trilayer \cite{Ge2021Control}. 
 To mention, magnetic moments of mirror-symmetric bands are controlled by the environment-induced surface energy, $\Delta_s$, and can be tuned across the range $g_v \sim 10\text{ to } d_v \sim 10^3$,  in contrast to the mirror-antisymmetric bands.
Similarly to rhombohedral graphite, individual Dirac points become almost indiscernible 
with a growing number of layers \cite{slizovskiy_films_2019}, the magnetic moment spreads over a broader range of momenta with the maximum $g_v \sim 100$ shifting away from the  $\boldsymbol{K}_\pm$ points.

\section{Conclusions}\label{Sec.V}

Overall, the presented study offers an effective model for thin films of rhombohedral graphite with a twin-boundary stacking fault. In particular, we show that all of such structures are semimetals, hence, they should feature ambipolar charge transport and  substantial magnetoresistance. In all these semimetallic systems, the Fermi level appears in the flat parts of the low-energy bands, hence, feature high density of states which potential to promote the formation of strongly correlated states of electrons.
Moreover, for several systems, such as 4ABA1 and a twinned 9-layer ABC film, 3ABA3,  we notice that the Fermi level is very close to the Lifshitz transitions in the low-energy bands spectrum. From thicker films,  we find that some bands have states localized at the twin boundary, which would make their flat band and possible correlated states less vulnerable to disorder effects form the environment.  

\section{Supporting information}
Supporting Information is available from the Wiley Online Library or from the author. 

\section{Acknowledgements}
This work was supported by EC-FET Core 3 European Graphene Flagship Project, EC-FET Quantum Flagship Project 2D-SIPC, EPSRC grants EP/S030719/1 and EP/V007033/1, and the Lloyd Register Foundation Nanotechnology Grant.

\section{Conflict of interest}
Authors declare no conflict of interest.

\newpage
\bibliography{Bibl}

\end{document}